\documentclass[10pt, aps, prd, amsmath, floats,floatfix,twocolumn, superscriptaddress]{revtex4-2}

\newcommand{\hh}{{\hspace{.3mm}}}

\def\sideremark#1{\ifvmode\leavevmode\fi\vadjust{\vbox to0pt{\vss
 \hbox to 0pt{\hskip\hsize\hskip1em

 \vbox{\hsize3cm\tiny\raggedright\pretolerance10000
 \noindent #1\hfill}\hss}\vbox to8pt{\vfil}\vss}}}%

\usepackage{mathrsfs}
\usepackage{natbib}
\usepackage{amsmath}
\usepackage{amssymb}
\usepackage{lmodern}

\usepackage[T1]{fontenc}
\usepackage[latin9]{inputenc}
\usepackage{geometry}
\usepackage{subfigure,lmodern, amsmath,amssymb, graphicx, pifont, adjustbox, bm, xcolor}
\usepackage{amsfonts}
\usepackage{geometry}
\usepackage{comment}
\usepackage{mathtools}
\usepackage{slashed}
\usepackage{ragged2e}

\usepackage{bm}

\newcounter{mnotecount}

\newcommand{\mnotex}[1]
{\protect{\stepcounter{mnotecount}}$^{\mbox{\footnotesize $\bullet$\themnotecount}}$ 
\marginpar{
\raggedright\tiny\em
$\!\!\!\!\!\!\,\bullet$\themnotecount: #1} }

\newcommand{\bg}{\boldsymbol{g}}

\newcommand{\otop}{\mathring{\top}}

\renewcommand{\=}{\stackrel{\Sigma}{=}}

\newcommand{\ce}{\mathcal{E}}
\newcommand{\ct}{\mathcal{T}}

\newcommand{\II}{{\rm  I\hspace{-.2mm}I}}
\newcommand{\IIo}{\hspace{0.4mm}\mathring{\rm{ I\hspace{-.2mm} I}}{\hspace{.0mm}}}

\newcommand{\IIIo}{{\mathring{{\bf\rm I\hspace{-.2mm} I \hspace{-.2mm} I}}{\hspace{.2mm}}}{}}

\newcommand{\IVo}{{\mathring{{\bf\rm I\hspace{-.2mm} V}}{\hspace{.2mm}}}{}}

\newcommand{\FFo}[1]{\mathring{\underline{\overline{\rm{#1}}}}}

\begin{document}

\title{Asymptopia of Kerr--de Sitter Black Holes}

\author{Samuel Blitz}
 \email{blitz@math.muni.cz}
  \affiliation{Department of Mathematics and Statistics, Masaryk University, Brno, Czech Republic}

\author{Jaros\l aw Kopi\'nski}
\email{jkopinski@ucdavis.edu}
  \affiliation{Center for Quantum Mathematics and Physics (QMAP), Department of Mathematics, University of California, Davis, CA, USA} 

\date{\today}
\begin{abstract}

\noindent
Exterior geometries of physical black holes are believed to asymptotically approach the Kerr--de Sitter spacetime at late times. A characteristic feature of that vacuum Einstein solution is the presence of a hidden symmetry generated by a closed conformal Killing--Yano tensor. 
Using this symmetry and modern conformal geometry technology, we find necessary conditions for generic solutions to asymptotically approach the Kerr--de Sitter metric.
Further, we constrain the admissible form of the geometric free data on the conformal infinity giving rise to this family of spacetimes and constrain it in terms of the stress-energy tensor. 
\end{abstract}

\maketitle

\noindent

\section{Introduction}
The Kerr spacetime is one of the most important exact solutions of the Einstein field equations because it describes an asymptotically flat rotating black hole. According to the ``establishment viewpoint'' of gravitational collapse~\cite{penr1}, any dynamical spacetime containing black holes and matter fields will eventually settle down to a state described by the Kerr black hole. Consequently, considerable effort has been devoted to understanding the global stability of this solution. Recently this problem has been solved for slowly rotating black holes in Minkowski and de Sitter spacetimes \cite{vasyhintz, fang, szeftel}. The latter case is particularly interesting from the perspective of cosmology and astrophysics, as observations indicate that our Universe has a positive cosmological constant. 
Therefore we ask what obstructs generic spacetimes from approaching the Kerr--de Sitter spacetime at late times?

The problem of characterizing both the initial data and the bulk geometry of general rotating black hole solutions of the Einstein field equations has been extensively analyzed. Invariant characterizations of such spacetimes include the Mars--Simon tensor \cite{mm0, simon, mm2}, concomitants of the Weyl tensor \cite{mm3, saez}, and generalized symmetries \cite{mm1,bk1, bk2, mp, ks}. We focus on symmetries generated by a closed conformal Killing--Yano (CKY) tensor. The most general solution of the vacuum Einstein field equations with cosmological constant admitting a closed CKY tensor is the Kerr--NUT--(anti-)de Sitter metric \cite{kubiznak1, houri}. This generalized symmetry underlies the complete integrability of the geodesic equation and the separability of the Hamilton--Jacobi, Klein--Gordon, and Dirac equations; see the review article~\cite{kubiznakliv} and the references therein.

Our approach leverages closed CKY tensors on the Kerr--de Sitter spacetime to characterize the far future configuration space of realistic black holes. To do so, we assume that the conformal infinity locally matches that of the Kerr--de Sitter black hole. In addition, both the conformal Einstein and CKY equations must be satisfied asymptotically. This constrains both the geometry and matter content of the black hole exterior. In particular, we derive a purely algebraic equation that restricts the free data for the conformal Einstein field equations on the conformal infinity in terms of the CKY tensor. This free data is then further restricted by the asymptotic matter content and vice versa.

\section{Background} \label{backg}
The Kerr--de Sitter metric in Boyer--Lindquist type coordinates $(t,r, \theta, \phi)$ is given by \cite{akcay}
\begin{equation} \label{kdsmet}
\begin{split}
\widetilde{g}_{KdS} = & - \tfrac{\Delta_r}{r^2 + a^2 \cos^2 \theta} \left(\tfrac{{\rm d} t - a \sin^2 \theta {\rm d} \phi}{1+\lambda}\right)^2 \\
& + \sin^2 \theta \tfrac{1 + \lambda \cos^2 \theta}{r^2 + a^2 \cos^2 \theta} \left(\tfrac{a {\rm d} t - (r^2 +a^2) {\rm d} \phi}{1+ \lambda}\right)^2 \\
& + (r^2 + a^2 \cos^2 \theta) \left( \tfrac{{\rm d} r^2}{\Delta_r} + \tfrac{{\rm d} \theta^2}{1+ \lambda\cos^2 \theta} \right),
\end{split}
\end{equation}
where $a$ is the rotation parameter, $\lambda := \frac{a^2 \Lambda}{3}$ with cosmological constant $\Lambda>0$, $\Delta_r := -\frac{\Lambda}{3} r^4 + (1 - \lambda) r^2 - 2 m r + a^2$, and $m$ is the mass. We have set the NUT charge to zero as current astrophysical observations of gravitomagnetic monopoles are not definitive, see e.g.  \cite{gravmon, gravmon2}. 

\smallskip
The metric \eqref{kdsmet} possesses a generalized symmetry generated by a closed 2-form $Q_{ab}$ solving the CKY equation~\cite{kubiznak1},
\begin{equation} \label{cky}
\nabla_a Q_{bc} - \nabla_{[a} Q_{bc]} + \tfrac{2}{3} g_{a[b} \nabla^d Q_{c]d} =  0\,.
\end{equation}
As the name implies, the above condition is conformally invariant under the following metric and $2$-form rescalings, 
$$
 g  \mapsto  \Omega^2 g\,, \quad Q \mapsto  \Omega^3 Q\,.
$$
This naturally leads to the \textit{CKY operator} mapping $Q_{ab}$ to $\mathcal{Q}_{abc}$, where
\begin{equation} \label{qtoq}
 \mathcal{Q}_{abc} :=\nabla_a Q_{bc} - \nabla_{[a} Q_{bc]} + \tfrac{2}{3} g_{a[b} \nabla^d Q_{c]d} \, ,
\end{equation}
so that the CKY Equation~\eqref{cky} becomes $\mathcal{Q}_{abc} =0$. Note that $\mathcal{Q}_{abc}$ has the following ``Cotton tensor type'' symmetries,
$$
\mathcal{Q}_{abc} = \mathcal{Q}_{a[bc]}\,, \quad \mathcal{Q}_{a}{}^a{}_b = 0\,, \quad  \mathcal{Q}_{[abc]} =0\,. 
$$
The Cotton tensor $C_{abc}:= \nabla_{b} P_{ca} - \nabla_c P_{ba}$ is the covariant curl of the Schouten tensor $P_{ab}:= \tfrac12 (R_{ab} - \tfrac{R}{6} g_{ab})$, where $R_{ab}$ is the Ricci tensor and $R$ the scalar curvature.

\smallskip
A solution of the closed CKY equation on the Kerr--de Sitter background is
\begin{equation} \label{qkds}
\begin{split}
Q_{KdS}   & = r {\rm d} t \wedge  {\rm d} r - a^2 \sin \theta \cos \theta {\rm d} t \wedge {\rm d} \theta  \\ & + a r \sin^2 \theta {\rm d} r \wedge {\rm d} \phi  \\ & +  a (r^2+a^2) \sin \theta \cos \theta {\rm d} \theta  \wedge {\rm d} \phi \,  .
\end{split}
\end{equation} 

The Kerr-de Sitter future asymptotic region is given by the limit $r \to \infty$. Introducing a coordinate $\rho = r^{-1}$ and conformal factor $\sqrt{\tfrac{3}{\Lambda}} \rho$ yields the compactified unphysical metric $g_{KdS} := \tfrac{3}{\Lambda} \rho^2 \widetilde{g}_{KdS}$. The intrinsic geometry of the conformal boundary $\Sigma$ is now determined by the conformally flat metric $\overline{g}$ induced by $g_{KdS}$ along the $\rho =0$ hypersurface \cite{bonga1},
\begin{equation} \label{gscri}
\overline{g} =  \tfrac{ {\rm d} t^2}{(1+\lambda)^2} - \tfrac{2 a  \sin^2 \theta {\rm d} t {\rm d} \phi }{(1+\lambda)^2}    + \tfrac{3 {\rm d} \theta^2}{\Lambda + \Lambda \lambda \cos^2 \theta} + \tfrac{3 \sin^2  \theta \rm{d} \phi^2}{\Lambda+\Lambda \lambda}\,.
\end{equation}
Since the normal vector $g_{KdS}^{-1}({\rm d} \rho, \cdot)|_{\Sigma}$ is timelike, the hypersurface $\Sigma$ is spacelike. This is a generic feature of asymptotically ($\Lambda>0$)-vacuum spacetimes.

\smallskip
 Importantly, the above Kerr--de Sitter compactification is not unique as the physical metric $\widetilde{g}_{KdS}$ induces many equivalent defining function--unphysical metric pairs~\cite{fndf}. These can be repackaged as scalar and tensor-valued conformal densities $\tilde{\sigma}$ and $\bg_{ab}$ such that
 \begin{equation} \label{kdsscale}
 \widetilde{g}_{KdS} = \tilde{\sigma}^{-2} \ \bg.
\end{equation}
Indeed, given a manifold $M$ equipped with a conformal class of metrics, a \emph{(scalar) conformal density} of weight $w$ is a section of $\mathcal{E}M[w]:= \big[(\wedge^4 TM)^2]^{\frac w{8}}$ or in other words is an appropriate power of a volume form. Weighted tensorial densities can be defined as sections of $\mathcal{B} \otimes \mathcal{E}M[w]$, where $\mathcal{B}$ is the corresponding tensor bundle. Hence, as seen in Equation~\eqref{kdsscale}, the Kerr-de Sitter metric induces a conformal class of metrics $\bg_{ab}$ (everywhere, including the conformal infinity) and a weight-1 density positive in the interior; positive weight-1 densities are called \textit{scales}. The interior scale $\tilde{\sigma}$ is special because even though it vanishes along the boundary, its exterior derivative does not. Scales with this property are called \textit{defining densities}. Finally note that the CKY tensor $Q$ and the image of a CKY operator $\mathcal{Q}$ are sections of $\wedge^2 T^* M[3]$ and $\otimes_{\circ}^3 T^* M[3]$ respectively, where $\circ$ denotes trace-free tensors.

The conformal boundary $\Sigma$  can now be viewed as a hypersurface embedded in the conformal extension of an (asymptotically) Kerr--de Sitter spacetime whose properties can then be understood by relating the intrinsic and extrinsic \emph{conformal} geometry of the embedding to the stress-energy tensor. 

\emph{Conformal fundamental forms} are weighted, rank~2 tensors that generalize extrinsic hypersurface curvatures to conformally embedded hypersurfaces~\cite{cff}. They are related to the asymptotic stress-energy tensor via the Einstein field equations~\cite {gk}. The first conformal fundamental form is the induced conformal class of metrics $\overline{\bg}$ on $\Sigma$. For 4-dimensional conformal manifolds, the second through fourth fundamental forms are of interest because they capture the initial data for the conformal Einstein equations. They are given by the trace-free extrinsic curvature $\IIo_{ab}:= \II_{ab} - H \overline{g}_{ab}$, where $H$ is the mean curvature, 
$$
\IIIo_{ab} := W_{\hat{n} ab \hat{n}} - \IIo_{a}{}^c \IIo_{bc} +  \tfrac13 \overline{g}_{ab} \IIo_{cd} \IIo^{cd},
$$
and
$$
\IVo_{ab} :=  C^{\top}_{(a|\hat{n}|b)}  -\overline{\nabla}^c W^{\top}_{\hat{n} (ab) c} +  H W_{\hat{n} a  \hat{n} b}\,,
$$
where $\hat{n}$ is the timelike unit normal vector to the boundary $\Sigma$, $\overline{\nabla}$ the Levi-Civita connection of $\overline{g}$ and $\top$ denotes projection along $\Sigma$. The fourth conformal fundamental form $\IVo$ is of utmost importance in dimension 4. In Euclidean signature, it is the (locally undetermined) image of the Dirichlet-to-Neumann map for the Poincar\'e-Einstein problem \cite{BGKW,GrahamDN}. In Lorentzian signature, $\IVo$ is a divergence-free and trace-free tensor that is part of the free data for Friedrich's vacuum conformal Einstein field equations~\cite{fr1, fr2}. One can show that it is equal to the electric part of the rescaled Weyl tensor in that setting. When matter fields are present, the divergence of $\IVo$ no longer vanishes, but instead can be expressed in terms of the stress-energy tensor. Similar relations hold for conformal boundaries of regularized spacetimes with initial isotropic singularities \cite{gkw}. These underlie cosmological models that extend the causal structure beyond the Big Bang, such as conformal cyclic cosmology~\cite{ccc1, ccc2, ccc3, ccc4, ccc5}. 
\section{Asymptotic Generalized Symmetry} \label{agssec}

We now consider a 4-dimensional Lorentzian conformal manifold $(M, \bg)$ with spacelike boundary $\Sigma$ equipped with a flat conformal class of metrics with representative $\overline{g}$ given in Equation~\eqref{gscri}. Let $Q$ be any 2-form such that 
\begin{equation} \label{boucky}
Q|_{\Sigma} =  Q_{KdS}|_{\Sigma}\,,
\end{equation}
where $Q_{KdS}$ is defined in \eqref{qkds}. Note that we do not assume that $Q$ solves any equation a priori, but will derive conditions on the geometry $(M,\bg)$ by assuming that $Q$ solves the CKY equation asymptotically near the conformal boundary. This setup ensures that the spacetime arising from $(M, \bg)$ will be ``Kerr--de Sitter-like'' in the neighborhood of conformal infinity.

Consider an \emph{almost closed} operator, mapping $Q_{ab}$ to $\mathcal{C}_{abc}$, where
$$
\mathcal{C}_{abc} := \sigma \nabla_{[a} Q_{bc]} - 3 (\nabla_{[a} \sigma) Q_{bc]} \,,
$$
and $\sigma$ is a defining density for $\Sigma$. A condition $\mathcal{C}_{abc}=0$ is equivalent to $Q_{ab}$ being a closed $2-$form with respect to the metric $\sigma^{-2} \bg_{ab}$. 

The system of equations giving rise to the asymptotic closed CKY 2-form can now be written in a conformally invariant way,
\begin{align}
  \mathcal{Q}_{abc} & = \mathcal{O}(\sigma^{\ell+1})\,, \label{acky1} \\
    \mathcal{C}_{abc} & = \mathcal{O}(\sigma^{\ell + 2}) \label{aclosed1}\,.
\end{align}
where $\ell>0$. Equation \eqref{acky1} is independent of any choice of scale, while Equation \eqref{aclosed1} is coupled in a conformally invariant way to $\sigma$.

\smallskip
Having introduced a coupled conformally invariant system of equations defining an asymptotic generalized symmetry, we may now employ a manifestly conformally invariant tractor calculus; see~\cite{cg} for a pedagogical introduction. Key details are given below.

Any $d$-dimensional conformal manifold $(M,\bg)$ with a conformal class of metrics $\bg$ of signature $(p,q)$ may be equipped with a rank $d+2$ vector bundle $\ct M$ called the \textit{standard tractor bundle}. A tractor $V \in \Gamma(\ct M)$  is given by a triple 
$$V^A= (\tau,\mu^a,\nu)\,,$$ 
for a given choice of metric $g \in \bg$, where $\tau \in \Gamma(\ce M[1])$, $\mu^a \in \Gamma(TM[-1])$ and  $\nu \in \ \Gamma(\ce M[-1])$. It transforms covariantly under a conformal rescaling of the metric. There is a canonical bundle metric $ \ h_{AB} \hspace{0.2cm}$ of signature~$(p+1,q+1)$, defined by $V^2:=h(V,V) = \mu^a \mu_a + 2 \tau \nu$ that is preserved by the \textit{tractor connection} $\nabla^\ct$ that acts on $V^A$ as
$$\nabla^{\ct}_a V := \left(\nabla_a \tau - \mu, \nabla_a \mu^b + P_a{}^b \tau + \delta_a{}^b \nu, \nabla_a \nu - \mu^b P_{ab} \right)\,.
$$
The connection $\nabla^\ct$ can be upgraded to the so-called \textit{Thomas-D operator}, which maps tractors of weight $w$ to tractors of weight $w-1$ and is given by
$$
D_A := \left((d+2w-2)w, (d+2w-2)\nabla^{\ct}, -(\Delta^\ct + Jw) \right)\,,
$$
where $J:=P_a{}^a$. For example, a \textit{scale tractor} corresponding to a scale $\tau$ is defined as $$I_{\tau} := \tfrac{1}{d} D \tau\,.$$

Given any defining density $\sigma$ for an embedded hypersurface $\Sigma$, we may always improve it to a scale giving rise to an asymptotically constant scalar curvature metric~\cite{gwl}. From now on we will assume that such improvement has been made and (slightly abusing the notation) denote the result by $\sigma$ as well. This is done to simplify the construction of conformally invariant derivative operators on $\Sigma$.

Note that
$$I^A D_A \= (d+2w-2) (\nabla_{n} - w H)\,,$$
where $n := \mathrm{d} s$ and $s$ is the $g$-representative of $\sigma$. Hence it is clear that $I^A D_A \circ \sigma^{\ell} = \mathcal{O}(\sigma^{\ell-1})$ and we may use powers of $I^A D_A$ to construct conformally invariant normal derivative operators on $\Sigma$ acting on tractors (or scalars as in \cite{gwb}). Those operators can be used to derive constraints arising from Equations~(\ref{acky1}) and~(\ref{aclosed1}), but for that we must first map them to tractors. This can be done using an insertion procedure; see~\cite{cff} for explicit formulas for the insertion operator $q$. Denoting $\mathcal{Q}_{ABC} := q(\mathcal{Q}_{abc})$ as the insertion of $\mathcal{Q}_{abc}$ into a tractor, we can then define conformally invariant normal operators corresponding to various hypersurface projections of that tensor according to,
\begin{equation} \label{frog-mister}
\resizebox{.88\hsize}{!}{$
\begin{split}
 (\delta^{(k)} \mathcal{Q})_{abc} :=& \left[(\overline{q}^* \circ \top \circ I \cdot D^k) \mathcal{Q}_{ABC} \right]_{abc},\\
(\delta^{(k)}_2 \mathcal{Q})_{ac} :=& [(\overline{q}^* \circ \top \circ I^B \circ I \cdot D^k) \mathcal{Q}_{ABC}]_{ac}\,, \\
(\delta^{(k)}_{12} \mathcal{Q})_{c} :=& [(\overline{q}^*  \circ \top \circ  I^A I^B \circ I \cdot D^k) \mathcal{Q}_{ABC}]_{c}\,,
\end{split}
$}
\end{equation}
where $\overline{q}^*$ is the formal adjoint of the induced insertion operator $\overline{q}$ on the hypersurface $\Sigma$ (used to extract a tensorial slot of a tractor) and $\top$ acts on tractors such that $U^\top_A I^A|_{\Sigma} = 0$. Explicit computations show that in generic spacetime dimensions, the leading transverse order (number of normal derivatives) of all of the operators above is $k$. A similar construction may be applied to $\mathcal{C}_{abc}$; the constraints arising from the closed condition obtained that way will be discussed in Section \ref{aKdSsec}. Note that tractor insertion of conformal Killing equations for differential forms has been studied in \cite{JosefRod}, see also \cite{Josef}.

We will be particularly interested in the first two operators from (\ref{frog-mister}). After writing them in terms of the CKY 2-form $Q_{ab}$, the leading transverse order terms reduce to the lower ones using commutators of covariant derivatives, i.e.
\begin{equation} \label{k2cff}
\begin{split}
    \otop \nabla_{\hat{n}}^k \mathcal{Q}_{abc} & \propto \FFo{k+2}_{\hh a[b} Q_{c]_{\circ} \hat{n}}^\top + \cdots, \\
    \otop \hat{n}^c \nabla_{\hat{n}}^k \mathcal{Q}_{(ab)c} & \propto\FFo{k+2}_{\hh(a}{}^c Q^{\top}_{b)_{\circ}c} + \cdots\,.
\end{split}
\end{equation}
In the above $\nabla^k_{\hat{n}} := \hat{n}^{a_1} ... \hat{n}^{a_k} \nabla_{a_1}... \nabla_{a_k}$, $\propto$ denotes equality up to a non-vanishing constant, and $\FFo{k+2}$ denotes the $(k+2)$th conformal fundamental form (as long as it is defined~\cite{fncff}). This can be used to constrain $\FFo{k+2}$ in terms of the projections of the asymptotic CKY 2-form on the conformal boundary. However, the coefficients of those leading terms in \eqref{frog-mister} depend on the dimension $d$ and order $k$ and vanish in certain cases. In particular, all of them are 0 when $d=4$ and $k=2$. This limits the utility of these operators at and beyond that order \cite{fnconj}. Note that this problem arises precisely when the conformal fundamental form  appearing in \eqref{k2cff} is $\IVo$, the free data for the conformal Einstein field equations in four dimensions. 

Despite those higher-order problems, we may use Equations~\eqref{frog-mister} to construct explicit first-order normal derivative operators on $\mathcal{Q}$ when $d=4$. This yields the following four conformally invariant operators,
\begin{equation*}
\begin{split}
& 2  \IIIo_{a[b} Q^{\top}{}_{\hspace{-0.2cm} c] \hat{n}}+ \overline{g}_{a[b} \IIIo_{c]}{}^d  Q^{\top}_{d \hat{n} } +  \ldots\,, \\ \vspace{0.2cm}
&   Q^{\top}{}_{ \hspace{-0.2cm}(a}{}^c \IIIo_{b)c} + \ldots\,, \\ \vspace{0.2cm}
& \top \nabla_{\hat{n}}^2 Q_{ab} - 2 Q^{\top}{}_{ \hspace{-0.2cm} [a}{}^c \IIIo_{b]c} +\ldots\,,\\ \vspace{0.2cm}
& \top \hat{n}^b \nabla_{\hat{n}}^2 Q_{a b} + \IIIo_{a}{}^b Q^{\top}{}_{ \hspace{-0.2cm} b \hat{n}} + \ldots\,,
\end{split}
\end{equation*}
where the subleading terms denoted by $\ldots$ depend on~$\II$, projections of $Q$, and curvatures intrinsic to $\Sigma$. The second and third lines above come from the symmetric and antisymmetric parts of the two-index operator from Equations \eqref{frog-mister}.

Since we do not expect to have conformally invariant operators at second and higher transverse orders, we will instead use the fact that the $k$-th normal derivative of $\mathcal{Q}_{abc}$ is conformally invariant if all lower-order normal derivatives of that tensor vanish. Such operators will be referred to as \emph{conditionally} (conformally) invariant. We will now use the above constraints on the asymptotic closed CKY 2-form $Q_{ab}$ to study the geometry of asymptotically Kerr--de Sitter spacetime.
\section{Rotating Black Holes and Generalized Symmetries} \label{aKdSsec}
Normalizing the cosmological constant to $\Lambda= 3$ the Einstein field equations become,
\begin{equation*} \label{efes}
\widetilde{R}_{ab} - \tfrac12 \widetilde{g}_{ab} \widetilde{R} + 3 \widetilde{g}_{ab} = \widetilde{T}_{ab}\,.
\end{equation*}
Splitting them into trace-free and trace parts we arrive at a system of conformally invariant equations coupled to the scale $\tilde{\sigma}$ \cite{gk},
\begin{equation} \label{efestr}
\begin{split}
  \nabla_a \nabla_b \tilde{\sigma}  & + \tilde{\sigma} P_{ab}  -\tfrac14 \bg_{ab} (\Delta \tilde{\sigma} + J \tilde{\sigma}) = \tfrac{\tilde{\sigma}}{2} \mathring{  \widetilde{T \hh} }_{ab}\,, \\
& I^2_{\tilde{\sigma}} = - 1 + \tfrac{\tilde{\sigma}^2}{12} \widetilde{T}\,.
\end{split}
\end{equation}
where $\tilde{\sigma}$ is a defining density for the conformal boundary~$\Sigma$, $\widetilde{T}:=\widetilde{T}_a{}^a$, and the singular physical metric $\widetilde{g} = \tilde{\sigma}^{-2} \bg$. We want to focus on solutions to Equations~\eqref{efestr} asymptotic to the Kerr--de Sitter metric in a neighborhood of~$\Sigma$. To achieve this we assume that the stress-energy tensor~$\widetilde{T}_{ab}$ vanishes on $\Sigma$ and that there exists a solution of Equations \eqref{acky1} and \eqref{aclosed1} -- an asymptotic CKY $2-$form. It follows that,
\begin{align}
    \nabla_a \nabla_b \tilde{\sigma} + \tilde{\sigma} P_{ab} -\tfrac14 \bg_{ab} (\Delta \tilde{\sigma} + J \tilde{\sigma}) &= \mathcal{O}(\tilde{\sigma}^{\ell}), \label{aEconstr} \\
    (\nabla_a \tilde{\sigma})(\nabla^a \tilde{\sigma}) -\tfrac{\tilde{\sigma}}{2}  (\Delta \tilde{\sigma} + J \tilde{\sigma}) + 1 &= \mathcal{O}(\tilde{\sigma}^{\ell+1}), \label{ASCconstr} \\
    \nabla_a Q_{bc} - \nabla_{[a} Q_{bc]} + \tfrac{2}{3} \bg_{a[b} \nabla^d Q_{c]d} &= \mathcal{O}(\tilde{\sigma}^{\ell+1}), \label{CKYconstr} \\
    \tilde{\sigma} \nabla_{[a} Q_{bc]} - 3 (\nabla_{[a} \tilde{\sigma}) Q_{bc]} &= \mathcal{O}(\tilde{\sigma}^{\ell + 2}) \label{aCconstr}\, ,
\end{align}
where $\ell \in \mathbb{N}_{>0}$. The specific powers of $\tilde{\sigma}$ in the right-hand sides above are required for the system to determine $\tilde{\sigma}$ and $Q$ to order $\tilde{\sigma}^{\ell+2}$. The first two equations define an asymptotically de Sitter spacetime, while the last two equip it with an asymptotically closed CKY 2-form. We call a solution to \eqref{aEconstr}-\eqref{aCconstr} an \emph{asymptotically Kerr--de Sitter spacetime} of order $\ell$ if the induced conformal class of metrics on $\Sigma$ is flat and $Q|_{\Sigma} =Q_{KdS}|_{\Sigma}$, as in Equation~\eqref{boucky}. Our goal will be to describe asymptotically Kerr-de Sitter spacetimes of order $2$, corresponding to the slowest degree of fall-off of matter fields as one approaches the conformal boundary. This will be done by firstly deriving purely geometric constraints, and then using them to restrict the leading behavior of the matter content of the spacetime.

\subsection{Solving Geometric Constraints}

An asymptotically de Sitter spacetime solving Equations~(\ref{aEconstr}) and~(\ref{ASCconstr}) with $\ell = 2$ necessarily has an umbilic conformal boundary $\Sigma$ with vanishing third conformal fundamental form \cite{gk}, i.e.
$$\IIo \= 0 \= \IIIo \,,$$
where ${\=}$ denotes equality on $\Sigma$. Proceeding order-by-order, we now derive further constraints arising from Equations~(\ref{CKYconstr}) and~(\ref{aCconstr}). The latter implies that the projected part of the asymptotic CKY tensor vanishes along the conformal infinity, i.e.
$$Q^\top_{ab} \= 0\,.$$
This is consistent with the corresponding solution on the compactified Kerr--de Sitter background given by~\eqref{qkds}. Equation~(\ref{CKYconstr}) now reduces to a set of constraints,
\begin{equation} \label{ckyeq0}
\begin{split}
    (\nabla_{\hat n} Q_{ab})^\top \=& - \overline{\nabla}_{[a} Q_{\hat{n}b]}^\top\,, \quad  \overline{\nabla}_{(a} Q_{b)_{\circ} \hat{n}}^\top \= 0\,, \\
   & \hat{n}^b \overline{g}_a{}^c \nabla_{\hat n} Q_{bc} \= H Q_{\hat{n} a}^\top \,.
\end{split}
\end{equation}
General manifestly conformally invariant constraint equations could have been derived, had we not fixed $\IIo \= 0 \= Q^\top$.

We now take higher-order (conditionally) conformally invariant normal derivatives of \eqref{CKYconstr} and \eqref{aCconstr} and impose the constraints obtained from the previous orders (cf. Section \ref{agssec}). Differentiating~\eqref{CKYconstr} and~\eqref{aCconstr} in the normal direction  yields two new constraints,
\begin{align*}
    (\nabla_{\hat n}^2 Q_{ab})^\top \=&  \  2 Q^{\top}{}_{ \hspace{-0.2cm}  \hat{n} [a} \overline{\nabla}_{b]} H - H \overline{\nabla}_{[a} Q^{\top}_{|\hat{n}|b]}\,, \\
    \hat{n}^b \overline{g}_a{}^c \nabla_{\hat n}^2 Q_{bc} \=&  \ \overline{\Delta} Q_{\hat{n} a}^\top +(P_b{}^b -H^2) Q_{\hat{n}a}^\top\,.
\end{align*}
Upon differentiating twice in the normal direction, we arrive at equations relating $(\nabla_{\hat{n}}^3 Q_{ab})^\top|_{\Sigma}$ and $\hat{n}^b \overline{g}_a{}^c \nabla_{\hat n}^3 Q_{bc}|_{\Sigma}$ to the geometry of the conformal boundary and $Q|_{\Sigma}$, as well as an algebraic constraint for $\IVo$,
\begin{align}
   \label{ivoalgcons}  \  2\IVo_{a[b} Q^{\top}{}_{\hspace{-0.2cm} |\hat{n}| c]} + \overline{g}_{a[b} \IVo_{c]}{}^d Q_{\hat{n}d}^\top \= 0\,.
\end{align}
It should be noted that in the setting of asymptotically de Sitter spacetimes one can only derive a differential constraint relating a divergence of $\IVo$ to the stress-energy tensor, see e.g. \cite{gk}.

Finally, as we work with asymptotically Kerr--de Sitter spacetimes of order $\ell=2$, we can only differentiate the conformally rescaled closed condition once more. However, the obtained equations provide no additional information when earlier constraints are imposed. 

We now focus on the algebraic equation~\eqref{ivoalgcons}.
Given that $Q|_{\Sigma} = Q_{KdS}|_{\Sigma}$ (see Equation \eqref{qkds}) and the induced metric $\overline{g}$ on $\Sigma$ from Equation \eqref{gscri} we may solve it for $\IVo$, yielding
\begin{equation} \label{ivoex}
\begin{split}
    \IVo =& A \big[ \mathrm{d}t^2 - 2 a \sin^2 \theta \mathrm{d} t \mathrm{d} \phi - \tfrac{(1+a^2)^2}{2(1+a^2 \cos^2 \theta)} \mathrm{d} \theta^2 \\&-\tfrac{1}{2}(1+a^2 - 3 a^2 \sin^2 \theta ) \sin^2 \theta\mathrm{d} \phi^2 \big] \,,
\end{split}
\end{equation}
where $A$ is a function on $\Sigma$. Note that $A= \tfrac{2 m}{(1+a^2)^2}$ for the Kerr--de Sitter metric.

\medskip

A spacetime satisfying the constraints above is asymptotically Kerr--de Sitter in a neighborhood of~$\Sigma$. This allows us to construct an asymptotic Killing vector field from the divergence of the asymptotic CKY 2-form $Q_{ab}$~\cite{jj2}. To achieve that, we need the physical metric and the CKY tensor to orders $-1$ and $-2$ in the defining function, respectively. Given a metric representative of the conformal infinity, a Fefferman--Graham expansion~\cite{FGbook} provides an explicit form of an asymptotically vacuum Einstein spacetime metric using the Gaussian normal coordinate to the boundary; see e.g.~\cite{jj} for the derivation of the Fefferman-Graham form of the rotating black hole metric with negative cosmological constant. Hence, given~$\overline{g}$ from Equation~\eqref{gscri}, the physical metric can be written as,
$$\widetilde{g} = s^{-2} g_{FG} \,.$$
Here $s$ is the geodesic distance from $\Sigma$ and the unphysical metric $g_{FG}$ obeys,
\begin{equation} \label{fgmetric}
g_{FG} = -d s^2 + \overline{g} + s^2 \overline{P} + \mathcal{O}(s^3)\,,
\end{equation}
where $\overline{P}$ is the Schouten tensor of the boundary metric~$\overline g$.
Note that \eqref{fgmetric} can only be used to recover the exact Kerr--de Sitter solution after inserting appropriate Neumann data at cubic order in $s$ (see e.g.~\cite{Aneesh} for the form of that data). All higher-order terms in the Fefferman--Graham expansion are then determined by the boundary metric and the cubic order term. In the next section, we will constrain the trace-free part of the Neumann data for asymptotically Kerr--de Sitter spacetimes.

An asymptotic CKY tensor of $g_{FG}$ can be obtained from Equation \eqref{ckyeq0} with boundary data of $Q_{KdS}$ arising from Equation \eqref{qkds}. This gives
$$Q = \mathrm{d} s \wedge (\mathrm{d} t -  a \sin^2 \theta \mathrm{d} \phi) + a  s\cos \theta \sin \theta \mathrm{d} \theta \wedge \mathrm{d} \phi + \mathcal{O}(s^2)\,.$$
The corresponding 2-form $\widetilde{Q}$ in the physical spacetime can be obtained from the conformal rescaling $\widetilde{Q} = s^{-3} Q$. Evaluating the leading-order divergence of $\widetilde{Q}$, we recover a Killing vector generated by $\partial_t$ in the asymptotic region,
\begin{equation} \label{akill}
\widetilde{\nabla}_a \widetilde{Q}^{ab} = 3(1 + a^2)^2 \delta_t^b + \mathcal{O}(s)\,.
\end{equation}

\subsection{Matter}
We now constrain the matter content of asymptotically Kerr--de Sitter spacetimes of order 2. The Einstein field equations \eqref{efestr} in this setting can be written as,
\begin{align*}
    \nabla_a \nabla_b \tilde{\sigma} + \tilde{\sigma} P_{ab} -\tfrac14 \bg_{ab} (\Delta \tilde{\sigma} + J \tilde{\sigma}) = \tfrac{\tilde{\sigma}^2}{2} \mathring{\tau}_{ab}\,, \\
  (\nabla^a \tilde \sigma)(\nabla_a \tilde \sigma) -\tfrac{\tilde{\sigma}}{2}  (\Delta \tilde{\sigma} + J \tilde{\sigma}) + 1 = \tfrac{\tilde{\sigma}^3}{12} \tau\,,
\end{align*}
where $\tau_{ab} := \tilde{\sigma}^{-1} \widetilde{T}_{ab}$ is the unphysical stress-energy tensor and $\tau:=\tau_a{}^a$. The asymptotic form of such $\tau_{ab}$ is given by  (see e.g. \cite{bonga1}),
$$\tau_{ab} = - \hat{n}_a \hat{n}_b \tau + \mathcal{O}(\tilde{\sigma})\,. $$
We also have the following relation between $\tau_{ab}|_{\Sigma}$ and~$\IVo_{ab}$~\cite{gk},
\begin{align} \label{divIvo}
    \overline{\nabla}^b \IVo_{ab} \=   p_a - \tfrac{1}{3} \overline{\nabla}_a \tau\,,
\end{align}
where $p_a := \lim_{\tilde{\sigma} \to 0 } \tilde{\sigma}^{-1} \tau_{\hat{n}a}^\top$ is obviously well-defined. The fourth conformal fundamental form $\IVo$ is given by \eqref{ivoex}, so we can now use \eqref{divIvo} to constrain the free function $A$, yielding
\begin{equation}
\begin{split}
& (1+a^2)^2 \partial_t A \= p_t - \tfrac13 \partial_t \tau \,, \\
&(1+a^2)^2 \partial_{\theta} A  \=  \tfrac23 \partial_{\theta} \tau - 2 p_{\theta}\,, \\
& (1+a^2)^2 (\partial_{\phi} A + 3 a \sin^2 \theta \partial_t A)   \= \tfrac23 \partial_{\phi} \tau - 2 p_{\phi}\,.
\end{split}
\end{equation}
This system relates geometry of asymptotically Kerr--de Sitter spacetimes to the leading behavior of the stress-energy tensor. When matter fields are not present we obtain $A=\tfrac{c}{(1+a^2)^{2}}$, which corresponds to the exact Kerr--de Sitter metric with mass parameter $\tfrac{c}{2}$.

\subsection{Gravitational Charges}
We will now provide a partial interpretation the of function $A$ in the admissible free data $\IVo$ from \eqref{ivoex} for the asymptotically Kerr--de Sitter spacetimes. This will be done by expressing quasi-local charges on the conformal boundary introduced in \cite{bonga1} in terms of that function (see \cite{Aneesh} for comparison with alternative definitions of charges in this setting). A gravitational charge $C_{\xi}[S]$ associated with a conformal Killing vector $\xi^a$ and a surface $S$ embedded in the conformal boundary is defined as
\begin{equation*}
C_{\xi}[N] :=  \frac{1}{8 \pi} \oint \displaylimits_N \left( \IVo_{ab} +\tfrac{1}{3} \tau \overline{g}_{ab} \right) \xi^a \hat{r}^b \mathrm{d} N\,,
\end{equation*}
where $\hat{r}^a$ and $d N$ are the unit normal to $N$ in $\Sigma$ and the area element of that surface with respect to metric representative $\overline{g}$, respectively. A simple calculation shows that the quantity above is conformally invariant. 

The charges corresponding to (conformal) Killing vectors $\partial_t$ and $\partial_{\phi}$ on the conformal boundary of the exact Kerr--de Sitter metric are independent of the surface $S$ and are proportional to mass and angular momentum parameters. In the current setting, those two vectors are also infinitesimal generators of symmetries of the conformal boundary (where the first one arises from the divergence of the asymptotic CKY tensor via Equation~\eqref{akill}). Hence, we can treat the corresponding charges as definitions of quasi-local mass and angular momentum of asymptotically Kerr--de Sitter spacetime. In particular, for the $t=\mathrm{ const}$ spheres $S_t$ in the metric representative $\overline{g}$ from~\eqref{gscri} we arrive at 
$$C_{\partial_t}[S_t] = \frac{1}{8 \pi}\oint \displaylimits_{S^2} \left(A + \frac{\tau}{3(1+a^2)^2} \right)\, \mathrm{d}S^2$$
and
$$C_{\partial_\phi}[S_t] = - \frac{3 a }{16 \pi} \oint \displaylimits_{S^2} A \sin^2 \theta \, \mathrm{d}S^2$$
where the integrals on the right-hand sides are done over the unit $2-$sphere equipped with the standard round metric, i.e. $\mathrm{d}S^2 = \sin \theta \mathrm{d} \theta \mathrm{d} \phi$.

In the case of $A=\frac{2m}{(1+a^2)^2}$ and $\tau=0$ we recover Kerr-de Sitter charges of \cite{bonga1}. Our expressions can then be thought as generalizations of (quasi-local) mass and angular momentum to the current scenario, with the relation between the function $A$ determining free data and those physical quantities.

\section{Accelerated Black Holes}
The methods presented in this article can be naturally extended to accelerated Kerr--de Sitter spacetimes. This class of solutions is particularly important from the perspective of characterizing gravitational radiation with~$\Lambda>0$. Indeed, results of \cite{senalv} (building on~\cite{senalv2}) show that only accelerating black holes of type~D produce gravitational radiation detectable at infinity. Hence, acceleration turns Kerr--de Sitter black holes into one of the prototypical ``toy models'' of spacetimes exhibiting non-vanishing radiation at the conformal boundary.

An acceleration parameter can be introduced in the Kerr--de Sitter solution by a conformal rescaling, see e.g.~\cite{kubiznak_accel}. This implies that the resulting spacetime also admits a CKY 2-form, whose failure to be closed is proportional to that parameter. Hence, a similar analysis as presented in this work can be utilized to characterize such class of solutions. That scenario is more natural from the perspective of a conformal geometry, as the (non-closed) CKY equation is conformally invariant and can be coupled to the Einstein field equations to determine the structure of asymptotic Kerr--de Sitter black holes with acceleration.
\begin{acknowledgements}
 The authors would like to thank Rod Gover and Andrew Waldron for useful discussions. S.B. would like to acknowledge support by the Operational Programme Research Development and Education Project No. Cz.02.01.01/00/22-010/0007541. J.K. is thankful to the Masaryk University at Brno for welcoming him during the Conformal Geometry and Submanifolds workshop in 2025, where part of this manuscript was written.
\end{acknowledgements}


\begin{thebibliography}{10}
\bibitem{fndf}
 A function $\rho$ is said to be defining for a hypersurface $\Sigma$ if $\rho|_{\Sigma} =0$ and $\mathrm{d} \rho |_{\Sigma} \neq 0$.
\bibitem{fncff}
Conformal fundamental forms can be defined up to order 4 when $d=4$, up to order $d-1$ when $d\geq 6$ is even,  and $\frac{d+1}{2}$ when $d$ is odd \cite{Blitz-willmore,cff}.
\bibitem{fnconj}
We have checked that there is no conformally invariant second-order differential operator with leading transverse symbol $\otop \hat{n}^c\nabla_{\hat n}^2 \mathcal{Q}_{[ab]c}$ in arbitrary dimensions and conjecture that the same is true for other projections. A similar statement for conformally invariant bulk operators has been proved in \cite{govhir}.
\end{thebibliography}
\end{document}